# Demonstration of Electric Double Layer Gating under High Pressure by the Development of Field-Effect Diamond Anvil Cell

Shintaro Adachi[1,a], Ryo Matsumoto[1,2], Sayaka Yamamoto[1,2], Takafumi D. Yamamoto[1], Kensei Terashima[1], Yoshito Saito[1,2], Miren Esparza Echevarria[1], Pedro Baptista de Castro[1,2], Peng Song[1,2], Suguru Iwasaki[1], Hiroyuki Takeya[1], and Yoshihiko Takano[1,2]

[1]MANA, National Institute for Materials Science (NIMS), 1-2-1 Sengen, Tsukuba, Ibaraki 305-0047, Japan

[2]Graduate School of Pure and Applied Sciences, University of Tsukuba, 1-1-1 Tennodai, Tsukuba, Ibaraki 305-8577, Japan

[a]Author to whom correspondence should be addressed: ADACHI.Shintaro@nims.go.jp

We have developed an approach to control the carrier density in various material under high pressure by the combination of an electric double layer transistor (EDLT) with a diamond anvil cell (DAC). In this study, this "EDLT-DAC" was applied to a Bi thin film, and here we report the field-effect under high pressure in the material. Our EDLT-DAC is a promising device for exploring unknown physical phenomena such as high transition-temperature superconductivity (HTS).

Strategic engineering the band structure and as well as the position of the Fermi level play a crucial role in the exploration of unknown phenomena, such as HTS. Applying pressure and/or tuning the carrier density is known to be an effective way of controlling the electronic structure. The recent evolution of pressure generation and field-effect devices have provided an alternative route for controlling the electronic structure of a material other than chemical doping. Such devices have allowed us to reach even intrinsically unstable phases in materials.

Recently, a lot of surprising phenomena have been discovered by the induction of carriers into materials using an EDLT structure[1-16]. This structure consists of an electrolyte that is sandwiched between a gate electrode and a sample surface. When a voltage is applied between the gate and the source electrode of the sample, an electric double layer (EDL) is oriented at a very short distance at the interface between the electrolyte and the sample. This electric double layer works similar to a dielectric in a capacitor. In the case of EDL gating, since the thickness of the ionic layer is at the nanometer length scales, a large amount of charges can be induced near the sample surface. The following references highlights recent examples of field-induced superconductivity using an EDLT structure: $SrTiO_3$ with a low carrier concentration[8,9], an atomically flat ZrNCl film from a cleaved single crystal[10,11], and many other two-dimensional systems[12-16].

On the other hand, the application of high pressure enables us to reach unknown phenomena, such as HTS, that is inaccessible at ambient pressure. The recent discoveries of HTS in hydrides using DAC[17-21] are a typical example of materials search under high pressure. It is worthy to stress that, this occurrence of pressure-induced HTS in hydrides was predicted by theoretical calculations[22-26] based on the Bardeen-Cooper-Schrieffer theory. Such a theory-guided search is becoming a promising approach for the exploration of unknown superconducting phases. Another successful of such an approach is our group recent discoveries of pressure-induced superconductors by using our original DAC[27,28] coupled with a data-driven approach[29-31].

As discussed above, both the EDLT and the DAC have been recognized as predominant tools for tuning the physical properties in materials. If the combination of high-pressure generation by DAC and the electrical field-effect by EDLT is achieved, it will unlock us a vast area of search space for unknown physical phenomena, that might have remained accessible only on theoretical studies. In this paper, we report the development of a device that combines an EDLT structure into a DAC (hereafter called EDLT-DAC) for this task. We found that our previously fabricated DAC with a micro-scale boron-doped diamond electrodes and an insulating un-doped diamond[27,28] is suitable as the base of a pressure generator with electric field-effect function. Furthermore, we found that the ionic liquid N,N-diethyl-N-methyl-N-(2-methoxyethyl)ammonium bis(trifluorom-ethanesulfonyl)imide (DEME-TFSI) effectively works as a pressure medium for the EDLT-DAC and previous report on the pressure dependence of the expansivity of DEME-TFSI from ambient pressure to below 0.6 GPa[32] showing good correspondence to our experimental results.

As a sample to verify the characteristics of the fabricated EDLT-DAC, we chose bismuth (Bi) film since it is well known that there are extremely few electrons and holes in Bi when compared to conventional metals[33-35], and therefore it is expected to exhibit a prominent response upon application of the electric field and pressure.

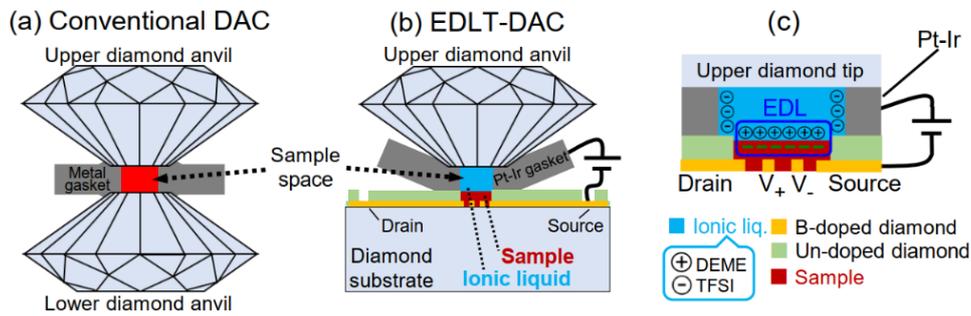

FIG. 1. (color online). Schematic image of a cross-section of (a) the conventional DAC and (b) the EDLT-DAC (this work). We used an ionic liquid (DEME-TFSI) as a pressure medium for the EDLT-DAC. In the EDLT-DAC, the gasket (Pt-Ir alloy) also works as a gate electrode. At the lower part of the EDLT-DAC, a boron-doped diamond was used as the electrodes, and the un-doped diamond was used as an insulating layer. (c) A cross-sectional image of a sample space of an EDLT-DAC while EDL gating is on.

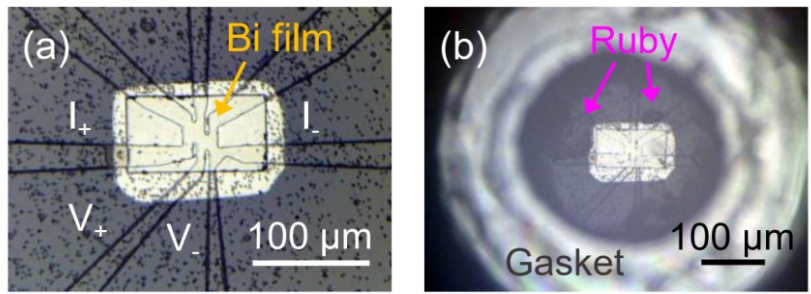

FIG. 2. (color online). (a) Optical microscope image of a Bi film on the lower diamond anvil. (b) Typical EDLT-DAC setup of a sample space with the ionic liquid (transparent in the figure), ruby crystalline powders, and Pt(80%)-Ir(20%) gasket. The diameter of the inner hole in the gasket was about 300 μm.

The schematic drawing of a conventional DAC[36,37] and the EDLT-DAC are shown in Figure 1(a) and 1(b). In a conventional DAC, high pressure is generated in the sample space, when two opposing diamond anvils with a polished flat tip (culet) are pressed. In our EDLT-DAC, a similar anvil with the culet diameter 600 μm was used as the upper anvil and our original diamond anvil with diamond electrodes[27,28] was used as the lower anvil. Metals are usually used as the sealing material (gasket) around the upper and lower anvils. The gasket in an EDLT-DAC needs to be electrochemically stable to work as the gate electrode. So we employed a Pt(80%)-Ir(20%) alloy as a gasket because it has electrochemical stability and enough hardness as a gasket since this alloy has almost the same Vickers hardness, HV (= 240)[38] as that of the SUS 316L[39,40]. The chosen Pt(80%)-Ir(20%) sheet with a thickness of 200 μm works not only as the gasket in the DAC but also as a gate electrode in the EDLT structure. We have put the sample and gasket on the lower anvil and then injected the ionic liquid DEME-TFSI into the sample space to construct the EDLT-DAC. Figure 1(c) shows a schematic image of the cross-section of the EDLT-DAC sample space. In this setup, when a positive voltage is applied between the Pt-Ir alloy gate and the source electrode, so that anions are attracted to the gate electrode while cations are attracted to the sample surface inducing electrons near the sample surface. The vicinity of the sample surface corresponds then, to the channel region of a conventional field-effect transistor.

Figure 2(a) shows an optical microscope image of the Bi film on the lower diamond anvil. The Bi thin film on the lower anvil was prepared by vapor deposition in a vacuum atmosphere and the thickness of the film was evaluated to be about 60－70 nm by atomic force microscopy. Figure 2(b) shows a top view of typical set-up of the sample space in the EDLT-DAC. The electrical resistance of the Bi film was measured using a standard four-probe method and the temperature dependence of electrical resistance (from 300 K to 2 K) showed a good agreement in the samples with the Bi films with a similar thickness reported in earlier studies[34,35]. The applied pressure was evaluated by using Piermarini's equation[41] as a function of a peak position of the $R_1$ ruby fluorescence line.

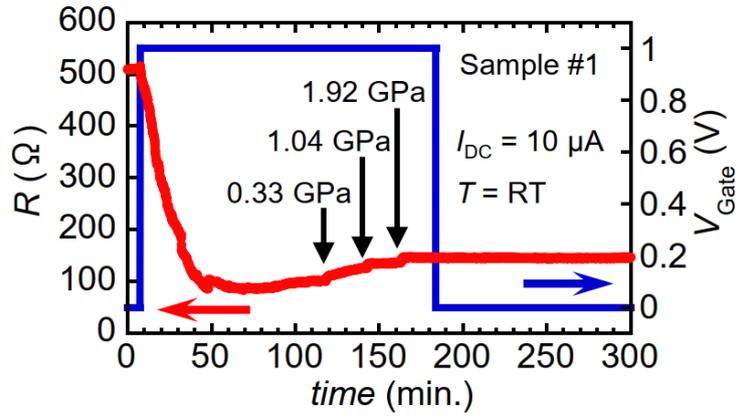

FIG. 3. (color online). The time dependence of the resistance in a Bi film at room temperature (RT) from ambient pressure up to 1.92 GPa. The black arrows indicate the point that pressure estimation was performed. Pressure estimation with EDL gating was carried out by a function of a peak position of the $R_1$ ruby fluorescence. Estimated pressures as of this time are described behind the black arrows.

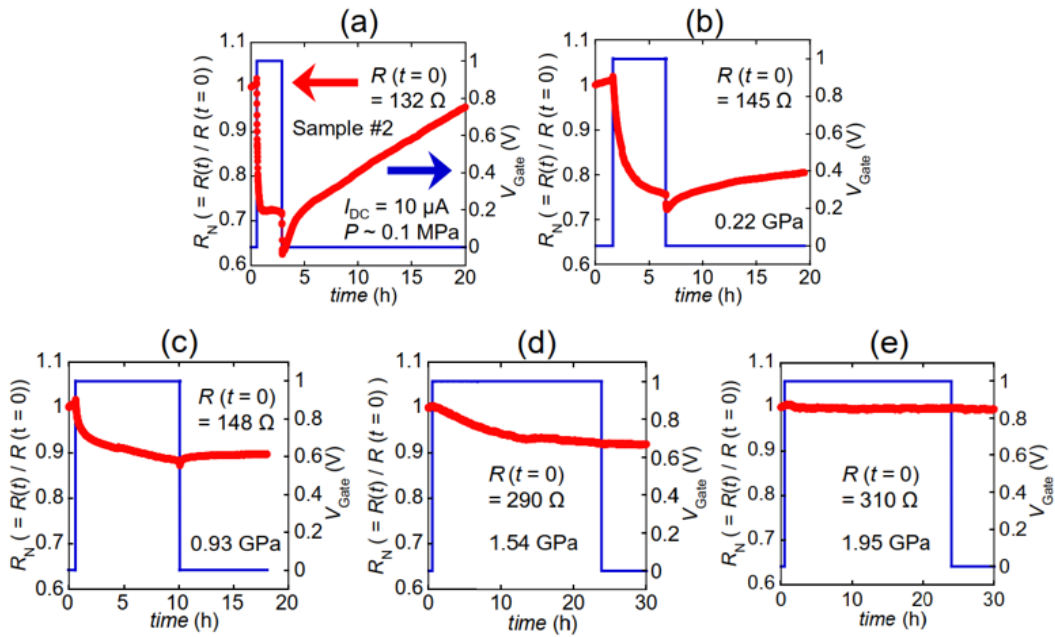

FIG. 4. (color online). (a-e) The time dependence of the normalized resistance of a Bi film at room temperature and that of the gate voltage (right axis) under various pressures from ambient pressure to 1.95 GPa. $R_0$ is the resistance value at $t = 0$.

Figure 3 shows the time dependence of the resistance of the Bi film at room temperature, where the right axis represents the applied gate voltage $V_G$ that was set either to be 0 V or 1 V. The leakage current flowing through the DEME-TFSI monitored by the signal of a digital multimeter our experiments. The leakage current never exceeded 20nA throughout the measurement, consistent with a previous work[32]. After setting $V_G$ = 1 V, the value of the resistance at ambient pressure decreased rapidly from about 510 Ω and saturated at 100 Ω in 40 minutes. This behavior indicates that carriers are induced in the channel of the Bi film by the EDL gating. After saturation, pressure was applied while keeping the gate voltage at 1 V, and the time at that each pressure was applied is indicated by the arrows with the corresponding pressure value. As we can see in Figure 3, the resistance slightly increases after each compressing. This may be due to the pressure effect on the Bi film or because of sample deformation. The most surprisingly, the resistance remains constant even after releasing the gate voltage. The stabilization of the sample resistance after releasing the gate voltage is similar to the results in the previous works where the EDL was stabilized by cooling (less than 190K[4]). Therefore, it seems that the compression of the ionic liquid would also stabilize EDL.

For a quantitative comparison of the time dependence of the resistance with EDL gating under various pressures, we show in Figure 4(a)-(e) the normalized resistance $R_N$ = $(R(t)/R_0)$ at room temperature, where $R_0$ indicates the initial resistance at $t$ = 0, for clarity, the range of the left axis is scaled from 0.6 to 1.1, and the right axis shows the value of the gate voltage. In the measurement at ambient pressure in Figure 4(a), after setting the gating voltage to 1 V, $R_N$ decreased rapidly from 1 to 0.72 and then $R_N$ value has become stable, showing a good correspondence to the results of Figure 3 at ambient pressure ($t$ < 115 min). After two hours, when we changed $V_G$ from 1 V to 0 V, $R_N$ showed a quick decrease. Then, a few minutes after, $R_N$ increased slowly and non-linearly. It should be noted that the behavior of resistance under ambient pressure (Fig. 4(a)) is in striking contrast to the one under high pressure (Fig. 3, $t \geq$ 184 min) where the constant resistance was observed even after $V_G$ was set to be zero.

Figure 4(b)-(e) shows the $R_N$ of Bi thin film in EDLT-DAC, under various pressures while $V_G$ was set to 1 or 0 V. By comparing Figures 4(a) and 4(b), we have found that two effects interfere with the field-effect when pressure is applied: (i) when a pressure of 0.22

GPa was applied, the resistance reduction rate due to the electric field effect was suppressed and (ii) by applying the same pressure of 0.22 GPa, the resistance recovery rate when $V_G$ was changed from 1 to 0 V was also suppressed. These two suppressing-effects have become more significant at higher pressure (0.93 GPa and 1.54 GPa, see Figures. 4(c) and 4(d))). Eventually at 1.95 GPa, the $R_N$ did not change even when the gating voltage was set to 1 V for 24 h.

Here we discuss the possible cause of the above effects on the resistance with EDL gating under high pressure, in relation to the EDL at low temperature. As an important characteristic of the ionic liquid, Yuan et al. reported that the ionic conductivity is decreased when the ionic liquid is cooled[4]. Such an interesting nature of the ionic liquid has drawn our attention, as it can be highly suitable for also maintaining the EDL under pressure. Namely, if pressure increased the viscosity of the ionic liquid DEME-TFSI, the EDL might be preserved. In general, it is known that the viscosity of a Newtonian fluid increases with increasing pressure: as it has been commonly observed for the organic liquid toluene[42], for an ionic liquid trihexyl(tetradecyl)phosphonium dicyanamide[43], and also for the glass forming liquids[44]. Moreover, ionic liquids are expected to undergo a glass transition at the point which their viscosity disappears due to applied pressure. A previous work[32] has reported the pressure dependence of the glass transition temperature of DEME-TFSI from 0.2 GPa up to 0.6 GPa. From these reported data, we deduced a linear relationship between the glass transition temperature ($T_G$) of DEME-TFSI and applied pressure $P$ (GPa): $T_G = 65(3)P + 195(1)$ from 0.2 GPa up to 0.6 GPa, where the number in parentheses are the error of the last digit. By extrapolating this linear equation to 300 K, we estimated the glass transition pressure $P_G$ to be 1.60(8) GPa $\leq P_G$ (T = 300K) $\leq$ 1.63(7) GPa. On the other hand, Our data of Figure 4(d) and Figure 4(e) suggests that the $P_G$ at room temperature would be 1.54 GPa $\leq P_G$ (T = RT) $\leq$ 1.92 GPa. Though the method of the current study is not the same as that of the previous work[32], we have found that the estimated $P_G$ value shows a good correspondence. Based on the current observation of the resistance of Bi thin film in the EDLT-DAC, it is implied that an electric double layer formed below $P_G$ may be stabilized by a glass-form ionic liquid caused by increasing pressure above $P_G$. It would be an interesting research topic in applied physics that by what condition the charge density of the EDL can be maintained even after the removal of electric field. More details about the stabilization of the electric double layer and also the applicability of the EDLT-DAC at higher pressure, lower temperature, and higher gate voltage, are needed to be investigated in future research.

In summary, we developed a field-effect diamond anvil cell (called the EDLT-DAC)

for tuning the properties of the materials and applied it to a thin film of Bi. We observed the electrical field-effect under high pressure in condensed matter. Also, we found that the EDL was stabilized by pressure. The stabilization-effect of the EDL under pressure may contribute to the development of devices such as transistors in the field of applied physics. We have demonstrated that our EDLT-DAC is capable of tuning the carrier density of the materials under high pressure, hence our developed device will certainly accelerate the exploration of physical phenomena that have not been accessible so far.

**Acknowledgment**

We are grateful to M. Fujioka, M. Tanaka, M. Nagao, H. Hara, Y. Sasama, K. Nakamura, A. Yamashita, S. Harada, T. Ishiyama, T. Yamaguchi, and T. Nojima for all the help. This work was supported by JSPS KAKENHI Grant No. JP17J05926, JP19H02177, JST-Mirai Program Grant No. JPMJMI17A2, and JST CREST Grant No. JPMJCR20Q4.